%% file: comm_main.tex
\documentclass[%
 preprint, 
nofootinbib,
 amsmath,amssymb,
 aps, physrev,
]{revtex4-2}
\usepackage{subfiles}
\usepackage{graphicx}
\usepackage{dcolumn}
\usepackage{bm}
\usepackage{hyperref}
\usepackage[version=4,arrows=pgf-filled,
textfontname=sffamily,
mathfontname=mathsf]{mhchem}

\usepackage{tikz}
\usepackage{quantikz}
\usepackage{algorithm}
\usepackage{algpseudocode}
\usepackage{placeins}
\makeatletter

\begin{document}

\title{Tackling instabilities of quantum Krylov subspace methods: an analysis of the numerical and statistical errors}%

\author{Maria Gabriela Jordão Oliveira}
 \email{maria.oliveira@nbi.ku.dk}
\affiliation{%
 NNF Quantum Computing Programme, Niels Bohr Institute, University of Copenhagen, Denmark
}%

\author{Karl Michael Ziems}
\affiliation{
School of Chemistry, University of Southampton, Highfield, Southampton SO17 1BJ, UK
}%

\author{Nina Glaser}
\email{ngl@chem.ku.dk}
\affiliation{%
 NNF Quantum Computing Programme, Niels Bohr Institute, University of Copenhagen, Denmark
}%
\affiliation{%
 Department of Chemistry, University of Copenhagen, Denmark
}%

\date{\today}

\begin{abstract}
Krylov subspace methods are among the most extensively studied early fault-tolerant quantum algorithms for estimating ground-state energies of quantum systems.
However, the rapid onset of ill-conditioning might make accurate energies difficult or even impossible to retrieve. 
In this communication, we analyse the numerical stability and statistical problems of these methods using numerical simulations both in the presence and absence of sampling noise. 
While in ideal numerical simulations the generalized eigenvalue problem indeed becomes unstable with increased Krylov subspace size, we find that, in realistic noisy settings, these methods do not primarily suffer from ill-conditioning.
Instead, statistical fluctuations dominate and can prevent reliable solution extraction unless appropriate regularization or filtering techniques are employed.
We consequently introduce two new metrics, the imaginary and unitary filters, that successfully assess the reliability of the obtained solutions without any knowledge of the true eigenspectrum.

\end{abstract}

\maketitle

\section{Introduction}

While quantum hardware has made tremendous progress in the past few years, current available noisy intermediate-scale quantum (NISQ) devices \cite{preskill_quantum_2018} remain far from fully fault-tolerant (FT) operation. 
Consequently, many quantum algorithms that promise to bring asymptotic advantages over classical methods cannot yet be executed on quantum devices for classically intractable problem instances. 
At the same time, the rapid development of hardware capabilities points towards the emergence of early-FT devices, which will offer limited fault-tolerant capabilities in the near future.
This outlook motivates the development of quantum algorithms capable of leveraging the capabilities of these emerging devices.  
As envisioned by Feynman \cite{feynman_simulating_1982}, a central application of quantum computing is the simulation of quantum systems. 
In this realm, Krylov subspace quantum methods \cite{parrish_quantum_2019, motta_determining_2020, cortes_quantum_2022,stair_multireference_2020,baker_block_2024, oumarou_molecular_2025,klymko_real-time_2022,kanasugi2025mirrorsubspacediagonalizationquantum, yeter-aydeniz_practical_2020, Zhang2023MeasurementefficientQK} are among the most explored ground-state energy estimation methods for early-FT devices. 
However, a commonly mentioned drawback of these methods is the ill-conditioning of the generalized eigenvalue problem (GEVP) spanned by the Krylov subspace as constructed from quantum circuit measurements \cite{kwao_generalized_2026, stair_multireference_2020,lee_efficient_2025, cortes_quantum_2022}.
Indeed, Ref.~\cite{lee_efficient_2025} states that due to ill-conditioning it is not yet clear if quantum Krylov methods will be superior to their classical counterparts in terms of runtime, and Ref.~\cite{cortes_quantum_2022} states that single-reference quantum Krylov methods may fail due to ill-conditioning for many problems of interest, including correlated molecular systems.
In this context, previous works have proposed strategies to mitigate ill-conditioning, ranging from the use of multiple initial references \cite{stair_multireference_2020} to elegant regularization of the problem \cite{epperly_theory_2022,lee_efficient_2025,lee_sampling_2024,Kirby_2024,Zhang2023MeasurementefficientQK}.
Ref.~\cite{epperly_theory_2022} also introduced two heuristic measures to assess the reliability of the solution, one based on the norm of the Ritz vector and one on the overlap between the initial state and the retrieved ground state (GS), which were, however, not found to be particularly reliable.

In this communication, we explore the numerical and statistical problems associated with quantum Krylov subspace methods, including the ill-conditioning problem and its practical implications in the finite shot limit.
We examine how different choices commonly made in quantum Krylov algorithms affect the condition number of the GEVP and how they impact the obtained energies. 
We further investigate different ill-condition mitigation techniques previously proposed in the literature, i.e., the use of a block of initial-references \cite{stair_multireference_2020} and the regularization of the overlap matrix via thresholding its singular values \cite{epperly_theory_2022}.
In addition, we introduce two simple diagnostic metrics to assess the reliability of the obtained solutions without requiring any knowledge of the true eigenstates.

\section{Results}

\subsection{Quantum Krylov algorithms theory}
Quantum Krylov algorithms rely on the idea of projecting the large Hamiltonian eigenvalue problem into a smaller, but educated, subspace, i.e., a Krylov subspace. 
Formally, given a block of $B$ initial references $\{|\psi^{(b)}_0\rangle\}_{b=1}^B$ and a operator $\hat{V}$, the Krylov subspace is defined as

\begin{multline}
{\mathcal{K}}_{(K+1)\times B}\left(\hat{V},\left\{|\psi^{(b)}_0\rangle\right\}_{b=1}^B\right)= \left\{|\psi^{(1)}_0\rangle, |\psi^{(2)}_0\rangle, ..., |\psi^{(B)}_0\rangle, \hat{V}|\psi^{(1)}_0\rangle, \hat{V}|\psi^{(2)}_0\rangle, \right. \\ \left. ...,\hat{V}|\psi^{(B)}_0\rangle, ..., \hat{V}^{K}|\psi^{(1)}_0\rangle,\hat{V}^{K}|\psi^{(2)}_0\rangle,...,\hat{V}^{K}|\psi^{(B)}_0\rangle \right\}.
\label{eq:krylov}
\end{multline}

Several choices for the subspace generator $\hat{V}$ have been proposed, including the Hamiltonian itself $\hat{H}$ \cite{baker_block_2024,oumarou_molecular_2025,Anderson_2025}, the real--time evolution operator $\hat{U}\left(t\right)= e^{-i\hat{H}t}$ \cite{parrish_quantum_2019,stair_multireference_2020,klymko_real-time_2022,cortes_quantum_2022,oliveira_quantum_2025,kanasugi2025mirrorsubspacediagonalizationquantum} and the imaginary--time evolution operator $\hat{U}\left(-it\right)= e^{-\hat{H}t}$ \cite{motta_determining_2020,yeter-aydeniz_practical_2020}. 
The use of $\hat{U}\left(t\right)$ relies on quantum circuits that are more amenable to early fault-tolerant devices, which is why we focus on this generator in this communication. 
It should be noted that regardless of the choice of $\hat{V}$, the orthogonality of the subspace is not guaranteed, and thus linear dependencies in the Krylov basis can arise, i.e, the problem can become ill-conditioned.

In Krylov subspace methods, the original Hamiltonian eigenvalue problem $\hat{H}\Psi=E\Psi$, where $\Psi$ is an eigenstate with energy $E$, is projected into the constructed subspace. This projection results in a generalized eigenvalue problem $\mathbf{T}\Phi=\Lambda \mathbf{S}\Phi$, where $\Phi$ are  eigenstates in the Krylov basis, and the eigenvalues $\Lambda$ can be trivially related to the original energies $E$. 
Here, $\mathbf{T}$ contains the expectation values of a function of $\hat{H}$, $f(\hat{H})$, given the elements of the basis, i.e., $T_{ij}=\bra{i}f(\hat{H})\ket{j}$ for $\ket{i}, \ket{j}\in{\mathcal{K}}_{(K+1)\times B}\left(\hat{U}\left(t\right),\left\{|\psi^{(b)}_0\rangle\right\}_{b=1}^B\right)$, and $\mathbf{S}$ is an overlap matrix with its elements being $S_{ij}=\langle i\ket{j}$. 
The function $f$ used to construct $\mathbf{T}$ is commonly either the identity, i.e.,  $T_{ij}=\bra{i}\hat{H}\ket{j}$, or the complex exponential, i.e., $T_{ij}=\bra{i} e^{-i\hat{H}\tau}\ket{j}$ with $\tau \in \Re\backslash\{0\}$.  
For simplicity, when $f$ is the identity, we refer to the algorithm as quantum block Krylov subspace H (QBKS-H), and when $f$ is the unitary complex exponential, we refer to it as QBKS-U.

\subsection{Solution reliability assessment}
Even for fully error-corrected quantum devices and in the limit of infinite sampling, the definition of the Krylov subspace in \autoref{eq:krylov} does not ensure orthogonality of its basis states, leading to (near) linear dependencies, and consequently to ill-conditioning of the GEVP. This manifests in a large condition number $\kappa(\mathbf{S})$, which quantifies the potential numerical instability of the classical diagonalization step, i.e., how sensitive the GEVP solution is to small perturbations in the input data. 
However, a large condition number does not by itself imply incorrect or unreliable results.
In fact, we show in this work that other measures can both filter spurious eigenvalues and assess the extent to which the problem was affected by errors. 
Moreover, even if the orthogonality of the basis was ensured and thereby ill-conditioning is circumvented alltogether, in practical implementations with a finite number of shots, sampling noise introduces errors in the matrix elements of $\mathbf{T}$ and $\mathbf{S}$, breaking the algebraic structure of the problem.

Specifically, in QBKS-H algorithm, the Hermiticity of $\hat{H}$ in principle guarantees that the lowest eigenvalue obtained from the GEVP, $\Lambda_0$, cannot lie below the true lowest eigenvalue of $\hat{H}$ through the Ritz variational principle \cite{RitzberEN,ArfkenWeber2005}.
The validity of this principle requires $\mathbf{T}$ to be Hermitian and $\mathbf{S}$ to be Hermitian and positive definite. 
However, when the $\mathbf{S}$ matrix is ill-conditioned or in the presence of noise, the positive definiteness might not be guaranteed, and so the Ritz variational principle may not hold anymore.
In practice, however, if $\mathbf{S}$ loses its positive definiteness, the eigenvalues $\Lambda$ may become complex.
Instead of simply disregarding the imaginary part of $\Lambda$, we here propose to utilize this additional information as the magnitude of the imaginary eigenvalue components can serve as an indicator of the error accrued in the calculation.
Specifically, we use the magnitude of the imaginary components of $\Lambda$ either as a metric of the reliability of the obtained eigenvalues or to filter out spurious solutions, which we hereafter refer to as imaginary filtering.
Given that our aim is to retrieve energies within chemical accuracy ($1$~Kcal/mol $\approx 1.6$~mHartree), i.e., $|\delta\Lambda_i|<1.6\times10^{-3}$ Hartree, the filtering criterion used for QBKS-H is $|\Im(\Lambda_i)|<1.6\times10^{-3}$, assuming that we tolerate errors of the same magnitude in both $E_i\approx\Re(\Lambda_i)$ and $\Im(\Lambda_i)$.
.

In QBKS-U algorithm, on the other hand, one solves the projected GEVP for the unitary (and generally non-Hermitian) time-evolution operator $\hat{U}(\tau)$ for a given time-step $\tau$.
As the Ritz variational principle only applies to Hermitian operators, the energies inferred from the time propagator are not variational.
Nevertheless, in the absence of errors, the GEVP eigenvalues should inherit the unitarity of the eigenvalues of $\hat{U}(\tau)$, and so, deviations of the eigenvalue norms from unity provide a direct indicator of the reliability of the obtained solutions, and can be used as a filtering criterion.
In this case, we propose the unitary filtering criterion as the measure of solution reliability, where we tolerate errors of the same magnitude in the norm $|1-|\Lambda_i||$ as the precision we target in the phase $ |\delta \text{arg}(\Lambda_i)|$.
Here, the eigenenergies $E_i$ are retrieved from the angles $\text{arg}(\Lambda_i)= -\frac{E_i\tau}{||\hat{H}||}$, and
given that for small phase errors $ \delta \text{arg}(\Lambda_i) \approx -\frac{\delta E_i\tau}{||\hat{H}||}$, we thus set
\begin{equation}
    |1-|\Lambda_0|| \sim |\delta \text{arg}(\Lambda_0)|<\frac{1.6\times10^{-3}\tau}{||\hat{H}||} \implies |1-|\Lambda_0|| <\frac{1.6\times10^{-3}\tau}{||\hat{H}||}.
\end{equation}

\subsection{Analysis}
For our numerical analysis, we choose two strongly correlated molecular structures in order to investigate the performance of the methodology on multi-configurational test systems that cannot be easily characterized by single-configuration methods, namely triangular \ce{BeH2} and rectangular \ce{H6}.
Further details of the numerical simulations and the reference state selection are given in Section~\ref{app:imp}.
To exclusively focus on the impact of numerical and statistical errors, we assume error-free implementations of the 
Hamiltonians $\hat{H}$ and the corresponding $\hat{U}$.
To assess the QBKS accuracy, we compare the retrieved GS energies with the GS energies obtained from exact diagonalization.

\begin{figure}
        \centering
        \includegraphics[width=1\linewidth]{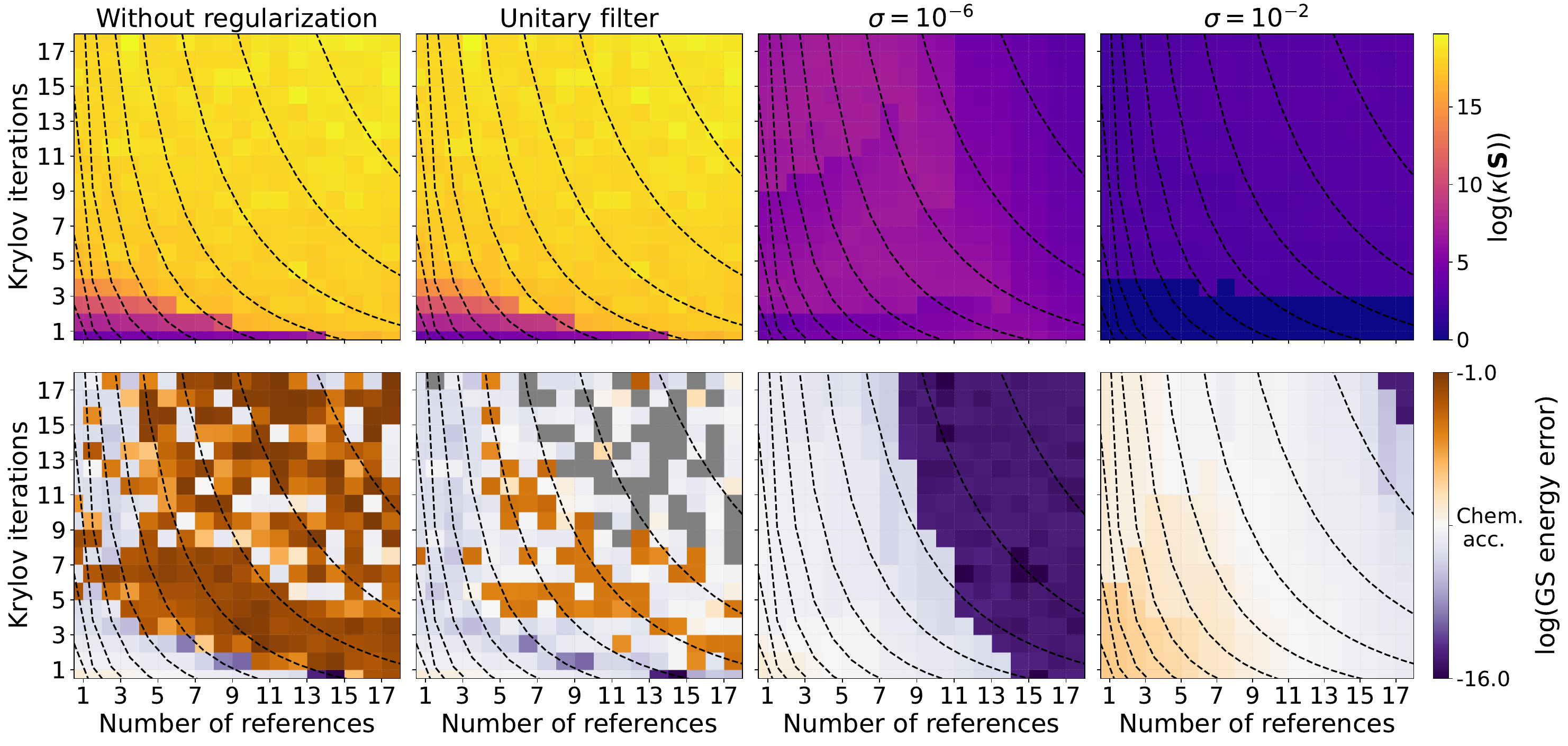}
        \caption{\textbf{QBKS-U algorithm convergence for \ce{BeH2} starting from different numbers of reference states and for different numbers of Krylov iterations.} The top panel shows the logarithm of $\kappa(\mathbf{S})$, whereas the bottom panel shows the logarithm of the absolute GS energy error. 
        The black dashed lines connect points with equivalent cost, i.e., the same number of distinct quantum circuits that are required to be sampled. The gray squares correspond to results where either the regularization procedure or the unitary filtering eliminated all the solutions.
        }
        \label{fig:size}
    \end{figure}

The dimension of the Krylov subspace can be increased in two ways: (1) by increasing the total number of Krylov iterations $K$; or (2) by increasing the number of initial reference states $B$. 
Therefore, to assess which approach is more effective to compute ground state (GS) energies, we show $\kappa(\mathbf{S})$ and the absolute GS energy error for different combinations of Krylov iterations and initial reference states for QBKS-U algorithm in \autoref{fig:size}.
The equicost lines connect points that require the measurement of the same number of distinct quantum circuits \cite{oliveira_quantum_2025}.

We compare the performance of the QBKS-U algorithm without regularization, both with and without unitary filtering, and at two different regularization levels in \autoref{fig:size}.
When applying the unitary filter, or, correspondingly, the imaginary filter for QBKS-H, solutions with norm deviations or imaginary components, respectively, exceeding the filtering thresholds are discarded.
The regularization strategy employed throughout this work follows Algorithm 2 of Ref.~\cite{oliveira_quantum_2025}, and relies on truncating all singular values of $\mathbf{S}$ smaller than a given threshold $\sigma$, which can be a fixed value or adaptively determined.
Further details of the employed regularization routine are given in Section \ref{app:imp}.

As can be seen in the top left panel of \autoref{fig:size}, more references lead to lower condition numbers for a given subspace size, as expected since the Krylov basis contains more linearly independent states by construction.
However, no corresponding improvement of the GS energy error is apparent for relatively small subspace sizes (bottom left corner), where near constant performance is observed even though there are large variations in the condition number of the unregularized overlap matrix (selected lineouts of the data are shown in Fig. 1 of the Supplementary Information (SI)).
Thus, $\kappa(\mathbf{S})$ as a metric by itself is not sufficient to assess the accuracy of the obtained GEVP solutions.
In addition, even though $\kappa(\mathbf{S})$ is smaller when using multiple references and few iterations, the computational cost to reach chemical accuracy tends to be smaller when using just a single reference and more iterations, as seen when following the equicost lines in the bottom left corners of the panels in \autoref{fig:size}.
The results of the QBKS-H algorithm yield similar conclusions, as do the rectangular \ce{H6} simulations for both QBKS algorithms (please see the SI for the corresponding figures).
Consequently, for the remainder of this work we focus on the single reference variants of the algorithms, which we denote by QKS-H and QKS-U, respectively.

Regarding the effects of the regularization procedure, the results in \autoref{fig:size} are consistent with the expectation that by removing smaller singular values of $\mathbf{S}$, and consequently some linear dependencies in the subspace, lower $\kappa(\mathbf{S})$ are obtained. However, this does not necessarily guarantee better energy results as the obtainable accuracy is highly sensitive to the threshold chosen.
When comparing the unregularized data (first column) with the results obtained after regularising the GEVP with a threshold of $\sigma=10^{-6}$ (third column), it is apparent that adequate regularization mitigates the ill-conditioning of the GEVP, increases the accuracy of the obtained energies, and also leads to smoother convergence with respect to the Krylov size.
However, high thresholds not only remove linear dependencies and numerical errors, but also significantly truncate the Krylov basis, which can lead to lower accuracy as seen for $\sigma=10^{-2}$ (fourth column).
The unitary filtering technique (second column), successfully post-removes the majority of erroneous GEVP solutions, thereby also improving the energy convergence, but without changing the subspace size before solving the GEVP or affecting $\kappa(\mathbf{S})$ as the GEVP is not modified by the filter.

\begin{figure}
        \centering
        \includegraphics[width=1\linewidth]{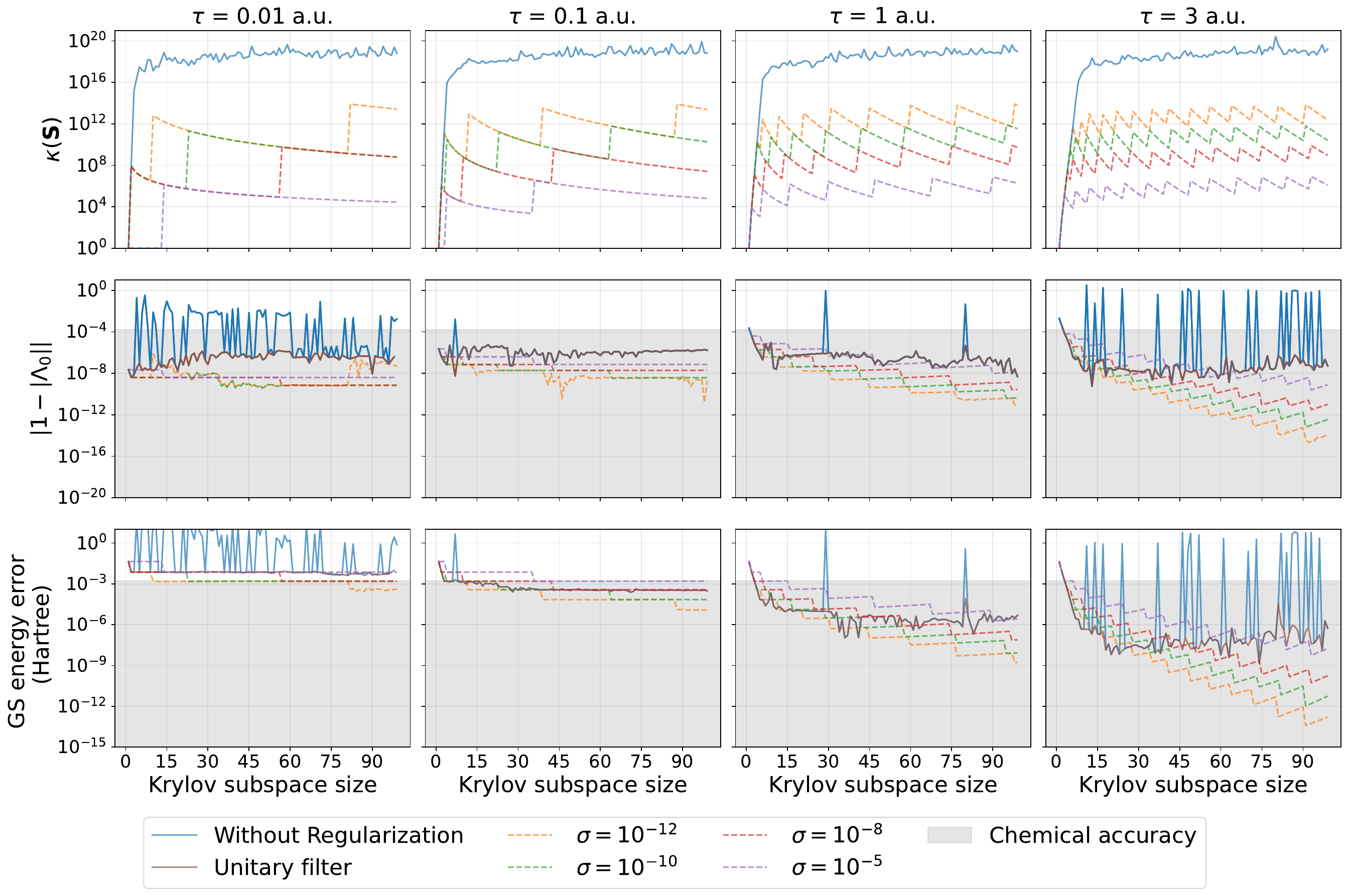}
        \caption{\textbf{QKS-U algorithm convergence for the \ce{BeH2} molecule using subspaces generated with different evolution times and regularization values.}
        The top panel shows the condition number $\kappa(\mathbf{S})$ in each Krylov iteration.
        The middle panel shows the error in the norm of the GS energy as a function of the Krylov iteration, whereas the bottom panel shows the absolute ground state energy error.
        }
        \label{fig:time}
    \end{figure}

Following, we analyse the impact of the time-step $\tau$ of the time evolution on the algorithm performance and numerical stability of the results.
Please note that, for QKS-U, $\tau$ should not be larger than $\pi$ a.u. to ensure that the eigenvalues of $\hat{H}$ can be unambiguously recovered from the eigenphases of $\hat{U}(\tau)$. 
The results for different $\tau$ values and several $\mathbf{S}$ regularization levels are shown in \autoref{fig:time}, where we report the absolute GS energy error, the deviation from unity of the lowest energy GEVP eigenvalue, and the condition number of the overlap matrix $\kappa(\mathbf{S})$.
In agreement with previous studies \cite{oliveira_quantum_2025,klymko_real-time_2022}, we find that the larger the time-step, the more accurate the retrieved GS energy for a given Krylov supspace size is.
For all the analysed $\tau$, the unregularized problem becomes ill-conditioned after a few Krylov iterations, with larger time-steps becoming unstable at slightly larger Krylov dimensions.
In the unregularized case, anomalous spikes in the energy correlate strongly with deviations from unitarity in the lowest-energy eigenvalue.
Thus, using our proposed unitary filtering, we can discard non-unitary eigenvalues, which significantly stabilizes the convergence by removing all spikes due to ill-conditioning (brown vs. blue lines in Fig.~\ref{fig:time}).
For longer time-steps and relatively small subspaces, the filtering leads to results that are equally or even more accurate than the best regularization threshold analysed.
In contrast, for larger Krylov dimensions the subspace becomes linearly dependent and regularizing it by removing the smallest singular values of $\mathbf{S}$ improves both convergence and accuracy, with lower thresholds performing better in the absence of noise.

Furthermore, when regularization is used, the step-wise increase in the condition number matches the corresponding decrease in the energy error.
Based on these results, while also taking into account that longer time evolutions are more costly to implement on quantum devices, we fix the evolution time to $\tau=1$ a.u. for QKS-U simulations, whereas we use $\tau=\frac{1}{||\hat{H}||}$ a.u. for QKS-H to ensure a consistent comparison.
Please note that the factor $\frac{1}{||\hat{H}||}$ in the latter arises because, unlike QKS-U, the QKS-H algorithm does not require Hamiltonian normalization.

\begin{figure}
        \centering
        \includegraphics[width=0.7\linewidth]{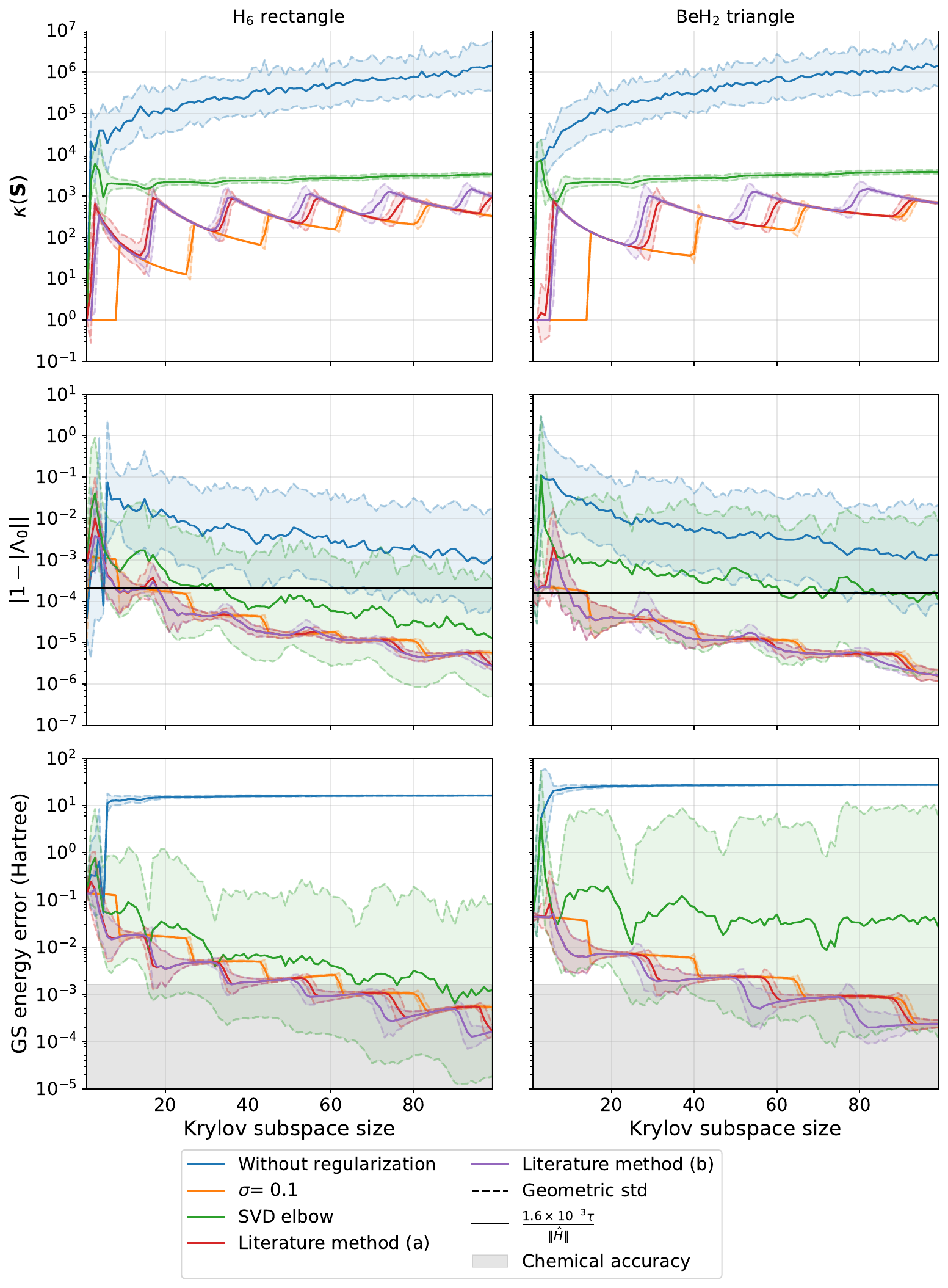}
        \caption{
        \textbf{QKS-U algorithm convergence for the \ce{H6} rectangle and the \ce{BeH2} triangle for different regularization schemes.} 
        The top panel shows $\kappa(\mathbf{S})$ in each Krylov iteration, the middle panel shows the error in the norm of the lowest energy GEVP solution, and the bottom panel shows the absolute error with regard to the exact eigenvalue.  
        The solid colour lines correspond to the geometric mean of 100 independent sampling runs with $10^{6}$ shots each per distinct quantum circuit, and the shaded area to the corresponding geometric standard deviation. The solid black line in the middle panel corresponds to the acceptance level of the norm error as employed in the unitary filtering.
        }
        \label{fig:Unoise}
    \end{figure}

\begin{figure}
        \centering
        \includegraphics[width=0.7\linewidth]{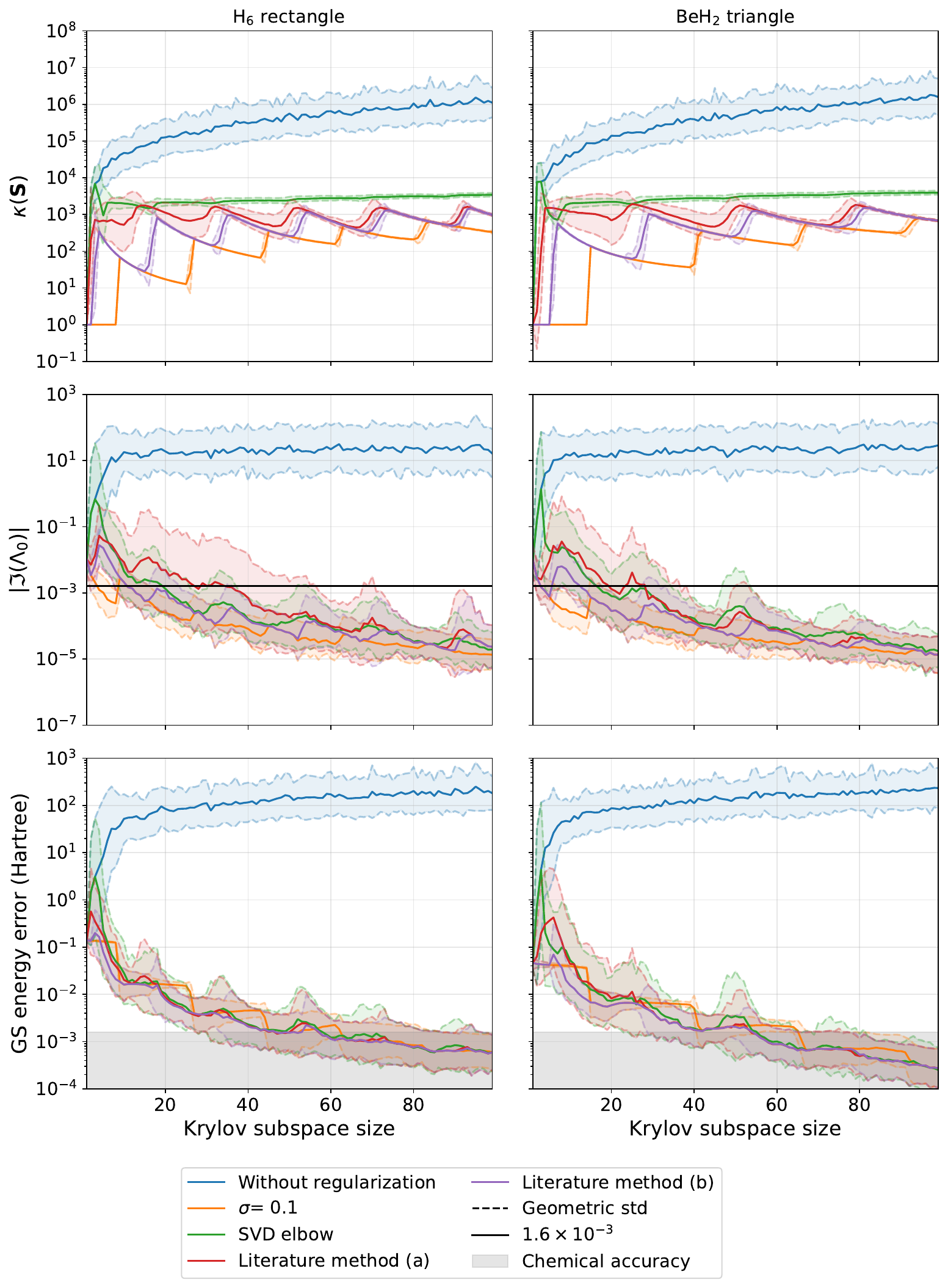}
        \caption{
        \textbf{QKS-H algorithm convergence for the \ce{H6} rectangle and \ce{BeH2} triangle molecules for different regularization schemes.}  
        The top panel shows $\kappa(\mathbf{S})$ in each Krylov iteration, the middle panel shows the error in the norm of the lowest energy GEVP solution, and the bottom panel shows the absolute error with regard to the exact eigenvalue.  
        The solid colour lines correspond to the geometric mean of 100 independent sampling runs with $10^{6}$ shots each per distinct quantum circuit, and the shaded area to the corresponding geometric standard deviation. The solid black line in the middle panel corresponds to the acceptance level of the error of the magnitude of $\Im(\Lambda_0)$ as employed in the imaginary filtering.
        }
        \label{fig:Hnoise}
    \end{figure}

While the condition number $\kappa(\mathbf{S})$ of noiseless simulations can be used to estimate how strongly sampling noise might affect the GEVP solution, we now explicitly assess the effect of sampling noise on the results.
To account for a finite number of quantum circuit executions, we next study the convergence of the QKS-U and QKS-H algorithms in the presence of shot noise in Figs.~\ref{fig:Unoise} and \ref{fig:Hnoise}, respectively.
For both algorithms, we compare several regularization strategies: (1) a fixed singular value threshold $\sigma$ as employed in the previous analysis; (2) discarding singular values below the “elbow” of the L-shaped $\log_{10}$ distribution of the singular values of $\mathbf{S}$ (see section I of the SI for an example), as identified using the \texttt{Kneed} Python package \cite{kneed}; (3) a noise- and iteration-dependent threshold labeled as method (a) corresponding to $\sigma = 0.1\times \text{Noise level} \times \text{Krylov size} \times N_F$, which is variation of the method proposed in Ref. \cite{Kirby_2024} to take into account an algorithm dependent normalisation factor, i.e,  $N_F=||\hat{H}||$ for QKS-U, and $N_F=\sqrt{\sum\limits_i |c_i|}$ , with $c_i$ being the coefficients in the Hamiltonian decomposition as a sum of Pauli strings, for QKS-H, respectively; 
and, the literature method (b) which uses $\sigma =2\times\sqrt{\text{Krylov size} \times \ln(2 \times \text{Krylov size})} \times \text{Noise level}$ as introduced in Ref.~\cite{lee_sampling_2024}.
Please note that regarding literature method (a), for the QKS-U algorithm, since the original spectrum is mapped to $[-1,1)$, this normalisation also impacts the overlap matrix $\mathbf{S}$, which needs to be accounted for, and so $N_F=||\hat{H}||$; For QKS-H, since the Pauli strings are sampled with a deterministic weighted sampling procedure \cite{arrasmith2020operatorsamplingshotfrugaloptimization}, the overall noise level depends on the sampling accuracy of the individual terms $H_i$ in the Hamiltonian decomposition $\hat{H}=\sum\limits_i c_i H_i$, and so $N_F=\sqrt{\sum\limits_i |c_i|}$.
Finally, in simulations with sampling noise, we use the proposed unitary and imaginary metrics to assess solution reliability rather than as filtering techniques, since preliminary results indicate that, although filtering smooths convergence and the metrics successfully flag erroneous solutions, they are by themselves not sufficient to achieve chemical accuracy without any further regularization.

In the presence of sampling noise, the $\kappa(\mathbf{S})$ values are lowered significantly compared to the noiseless case, indicating that the sampling noise alleviates linear dependencies. However, the limiting factor now becomes the sampling accuracy of the matrix elements. 
In this case, even though the problem is not ill-conditioned \textit{per se}, it is not possible to obtain accurate solutions without appropriate regularisation techniques due to the statistical noise present in the GEVP (blue lines in Figs.~\ref{fig:Unoise} and \ref{fig:Hnoise}).
In fact, the unregularized GS energy error not only exceeds chemical accuracy but is also clearly flagged as an erroneous solution by our unitary and imaginary filtering metrics, as indicated by the blue curves lying almost entirely above the horizontal black line in the middle panels of Figs.~\ref{fig:Unoise} and \ref{fig:Hnoise}, respectively.
For both QKS-U and -H, adequate regularization techniques enable the retrieval of the GS energy within chemical accuracy for both strongly correlated test systems.
The regularization with the elbow method (green lines) yields the least accurate average energies and the largest standard deviations. This behaviour is reflected by the largest deviations from unitarity and largest imaginary components, in the respective cases.
Regarding the other regularization techniques, although their overall performance is similar, the adaptive threshold as proposed in Ref.~\cite{lee_sampling_2024} (purple lines in Figs.~\ref{fig:Unoise} and \ref{fig:Hnoise}, labelled as literature method (b)) yields the highest accuracy for both algorithms. 
For the QKS-U algorithm, fixed-threshold and literature methods convergence steps occur at different Krylov subspace sizes, with the literature method (b) achieving chemical accuracy at smaller subspace dimensions for both molecules.
With the same overall shot budget per matrix element, QKS-U appears to result in moderately better mean accuracy and lower standard deviations than QKS-H with the given regularization techniques, albeit with the drawback of a less smooth step-wise energy convergence.

It should be noted that, although there is a clear difference in the performance of the different regularization methods, the resulting condition number values are of the same magnitude.
This further indicates that $\kappa(\mathbf{S})$ is not a suitable metric to assess the reliability of the results.
The newly proposed metrics $|1-|\Lambda_0||$ and $\Im(\Lambda_0)$, on the other hand, provide a much better indication of the (un)reliability of the obtained solutions (middle panel of \autoref{fig:Unoise} and \autoref{fig:Hnoise}), respectively.
We emphasize that the proposed metrics do not assess the accuracy of the obtained solution but rather its reliability, i.e., a value below the filtering threshold is not guaranteed to lie within chemical accuracy if the calculation has not yet converged, but an obtained GEVP result not accepted by the filtering is likely an erroneous solution.

\section{Discussion}

In this communication, we analyse the numerical and statistical issues commonly present in Krylov subspace methods.
We find that through the use of adequate regularisation and filtering techniques, Krylov subspace algorithms can yield chemically accurate results for strongly correlated systems even in the presence of sampling noise.
We also propose two simple metrics to assess the reliability of the retrieved solutions.

In the absence of sampling noise, the QKS-U and -H algorithms exhibit very similar convergence behaviour.
Longer evolution time steps lead to more accurate solutions, but appropriate regularization allows accurate results even with shorter time steps. 
Furthermore, the erroneous spikes in the retrieved energy due to numerical instabilities are clearly identified as unreliable solutions by the unitary and imaginary filtering, and when filtered out, the convergence becomes smooth as expected.
Regarding the expansion of the subspace, even though more initial references tend to lead to smaller condition numbers in the first few iterations, we find that the overhead of measuring additional quantum circuits makes a single reference with proper regularization the most cost-effective approach for obtaining the GS energy.

In the presence of shot noise, the condition number is lowered considerably compared to the noise-less case.
Even though in this setting the classical generalized eigenvalue problem is no longer ill-conditioned, the GS energy cannot be obtained without proper regularization to mitigate the statistical noise.
Also in this context, our two filtering metrics can serve as a reliability measure of the obtained solutions, without requiring any knowledge of the true GS energy.

Overall, we find that ill-conditioning does not prevent quantum Krylov methods from accurately finding the GS energy, as adequate regularization techniques succeed in removing the nearly linearly dependent elements of the basis.
Furthermore, in more realistic settings with sampling noise, the GEVP is not ill-conditioned but suffers from instabilities arising from the statistical fluctuations in the matrix elements, which can also be mitigated through regularization.
Therefore, while the condition number is a suitable measure to assess how sensitive the classical diagonalization of the GEVP is to variations in the input data, and is as such often used as a key quantity to gauge the stability of Krylov subspace methods, we find that the metrics proposed in this work provide a more practical measure of the reliability of the computed solutions.

\section{Methods}
\label{app:imp}
We employ two strongly correlated molecular structures as test systems in this work, namely triangular beryllium hydride (\ce{BeH2}) with a bond length of $1.3$~\AA~ and $\angle \ce{HBeH}$ of $60
^\circ$, as well as a rectangular system of 3 \ce{H2}-units (\ce{H6}) with intra-\ce{H2} bond lengths of $1.06$~\AA and inter-\ce{H2} distances of $1.59$~\AA. 
Both systems are represented in a molecular orbital basis obtained from a Hartree--Fock mean-field calculation in the STO-3G basis set \cite{basis}.
The fermionic Hamiltonians are mapped onto spin operators through the Jordan-Wigner transformation \cite{JW}, and for \ce{BeH2} the freeze core reduction is employed through Qiskit's \texttt{FreezeCoreTransformer} class \cite{qiskit2024}.

For the initial reference selection, we follow a scheme equivalent to the one in Ref.~\cite{stair_multireference_2020}.
If open-shell electronic configurations are present in the GS, they are grouped according to their spin occupation patterns, while closed-shell configurations are treated independently.
The resulting states are then ordered by their overlap with the true GS \cite{stair_multireference_2020}.
When a single reference is employed, the state resulting from the above scheme with the largest overlap with the true GS, i.e., the first state in the list of references, is utilized.

To regularize the GEVP, we first perform a singular value decomposition of the overlap matrix, $\mathbf{S} = \mathbf{W \Sigma Z}$, and the $\mathbf{S}$ and $\mathbf{T}$ matrices are then projected to the truncated subspace as
\begin{equation}
\mathbf{\tilde{S}} = \left( \mathbf{W}^{-1} \mathbf{S} \mathbf{Z}^{-1} \right)_{\mathcal{I}\mathcal{I}}, \qquad
\mathbf{\tilde{T}} = \left( \mathbf{W}^{-1} \mathbf{T} \mathbf{Z}^{-1} \right)_{\mathcal{I}\mathcal{I}} .
\end{equation}
where $\mathcal{I}$ is the index set with singular values above the truncation threshold, i.e., $\mathcal{I} = \{ i \mid \Sigma_i > \epsilon \}$.

For all shot noise simulations, we sample $2\times10^6$ shots per matrix element, i.e., $10^6$ samples each for the real and the imaginary part of each expectation value in $\mathbf{T}$ and $\mathbf{S}$.
To compute the Hamiltonian expectation values needed for QKS-H, we employ a deterministic weighted sampling according to Ref.~\cite{arrasmith2020operatorsamplingshotfrugaloptimization}.
The $10^6$ shots for computing the real or imaginary part are distributed among the Pauli strings $H_i$ in $\hat{H}$ depending on their weight, i.e., assuming that $\hat{H}=\sum\limits_i c_iH_i$, with $c_i$ being the coefficient associated with the Pauli string $H_i$, the number of samples used for each Pauli string $H_i$ is given by the closest integer to $\frac{|c_i|}{\sum\limits_j|c_j|} \times 10^6$.

\section*{Data availability}
The data presented in this work was generated using the code that the authors made openly available and is available upon reasonable request.

\section*{Code availability}
The data presented in this work was generated using the code made public available at \url{https://github.com/NQCP/NQCP-AA-pub-QBKSP_eigensolver}. 
\section*{Acknowledgements}
M.~G.~J.~O. and N.~G. gratefully acknowledge support by the Novo Nordisk Foundation, Grant number NNF22SA0081175, NNF Quantum Computing Programme.
\newpage

\bibliography{references}

\pagebreak
\widetext
\begin{center}
\textbf{\large Supplementary information for: Tackling instabilities of quantum Krylov subspace methods: an analysis of the numerical and statistical errors}
\end{center}
\FloatBarrier
\setcounter{equation}{0}
\setcounter{figure}{0}
\setcounter{table}{0}
\setcounter{page}{1}
\setcounter{section}{0}
\makeatletter
\renewcommand{\theequation}{S\arabic{equation}}
\renewcommand{\thefigure}{S\arabic{figure}}
\renewcommand{\bibnumfmt}[1]{[S#1]}
\renewcommand{\citenumfont}[1]{S#1}

\subfile{comm_si.tex}
\end{document}

%% file: comm_si.tex
\title{Supplementary information for: Tackling instabilities of quantum Krylov subspace methods: an analysis of the numerical and statistical errors}%

\author{Maria Gabriela Jordão Oliveira}
 \email{maria.oliveira@nbi.ku.dk}
\affiliation{%
 NNF Quantum Computing Programme, Niels Bohr Institute, University of Copenhagen, Denmark
}%

\author{Karl Michael Ziems}
\affiliation{
School of Chemistry, University of Southampton, Highfield, Southampton SO17 1BJ, UK
}%

\author{Nina Glaser}
\email{ngl@chem.ku.dk}
\affiliation{%
 NNF Quantum Computing Programme, Niels Bohr Institute, University of Copenhagen, Denmark
}%
\affiliation{%
 Department of Chemistry, University of Copenhagen, Denmark
}%

\date{\today}
\maketitle

\section{\ce{BeH2} supplementary results}
\begin{figure}
    \centering
    \includegraphics[width=1\linewidth]{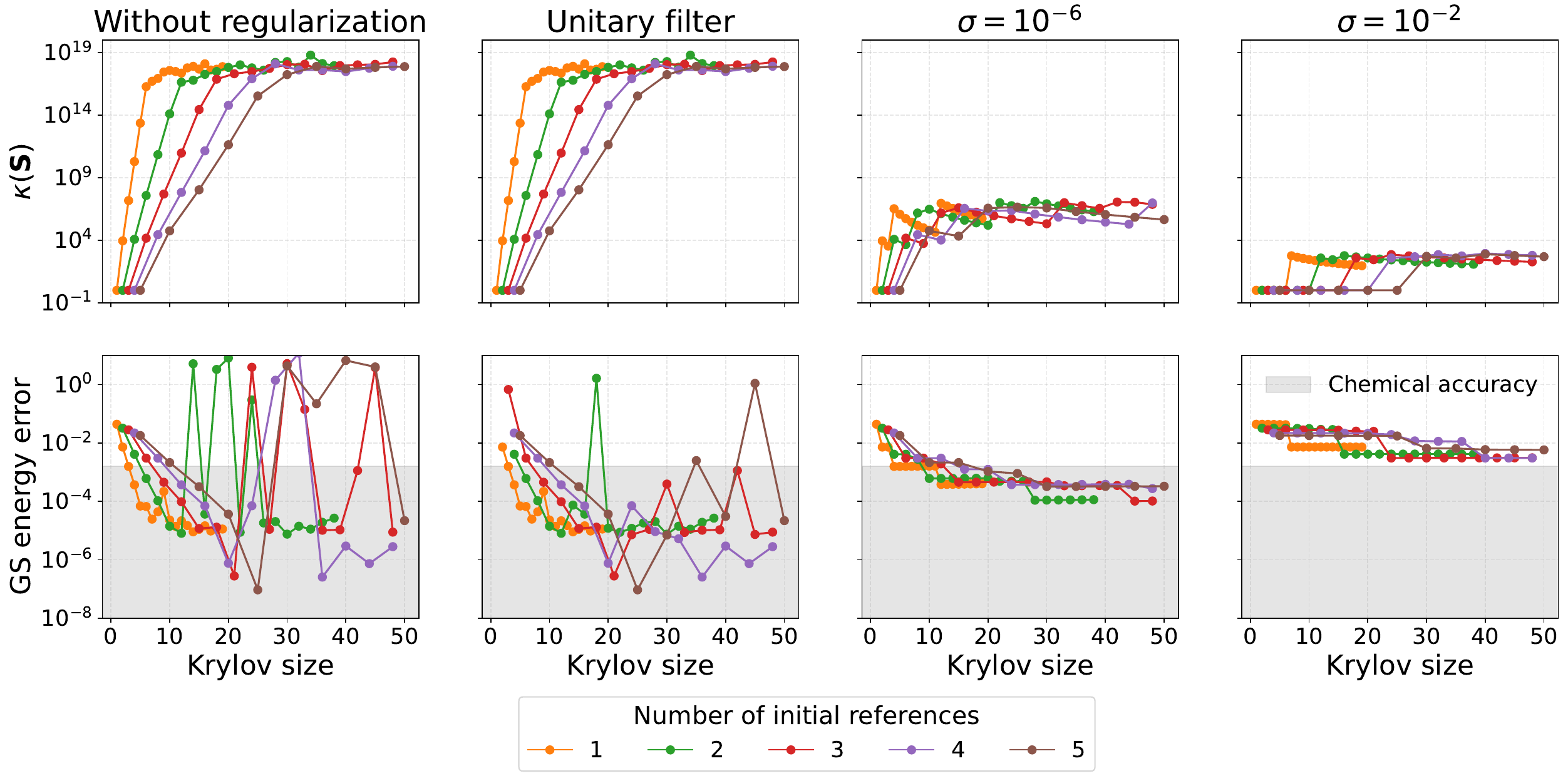}
    \caption{\textbf{QBKS-U algorithm convergence for the \ce{BeH2} triangle starting from different numbers of reference states and for different Krylov subspace sizes.}
    The top panel shows the logarithm of the condition number, whereas the bottom panel shows the logarithm of the absolute GS energy error with regard to the exact eigenvalue.}
    \label{fig:sizeandref}
\end{figure}

This section contains the supplementary Q(B)KS-U and -H algorithms analysis. In \autoref{fig:sizeandref}, the Krylov subspace growth analysis for the QBKS-U algorithm is alternatively visualised.

\begin{figure}
        \centering
        \includegraphics[width=1\linewidth]{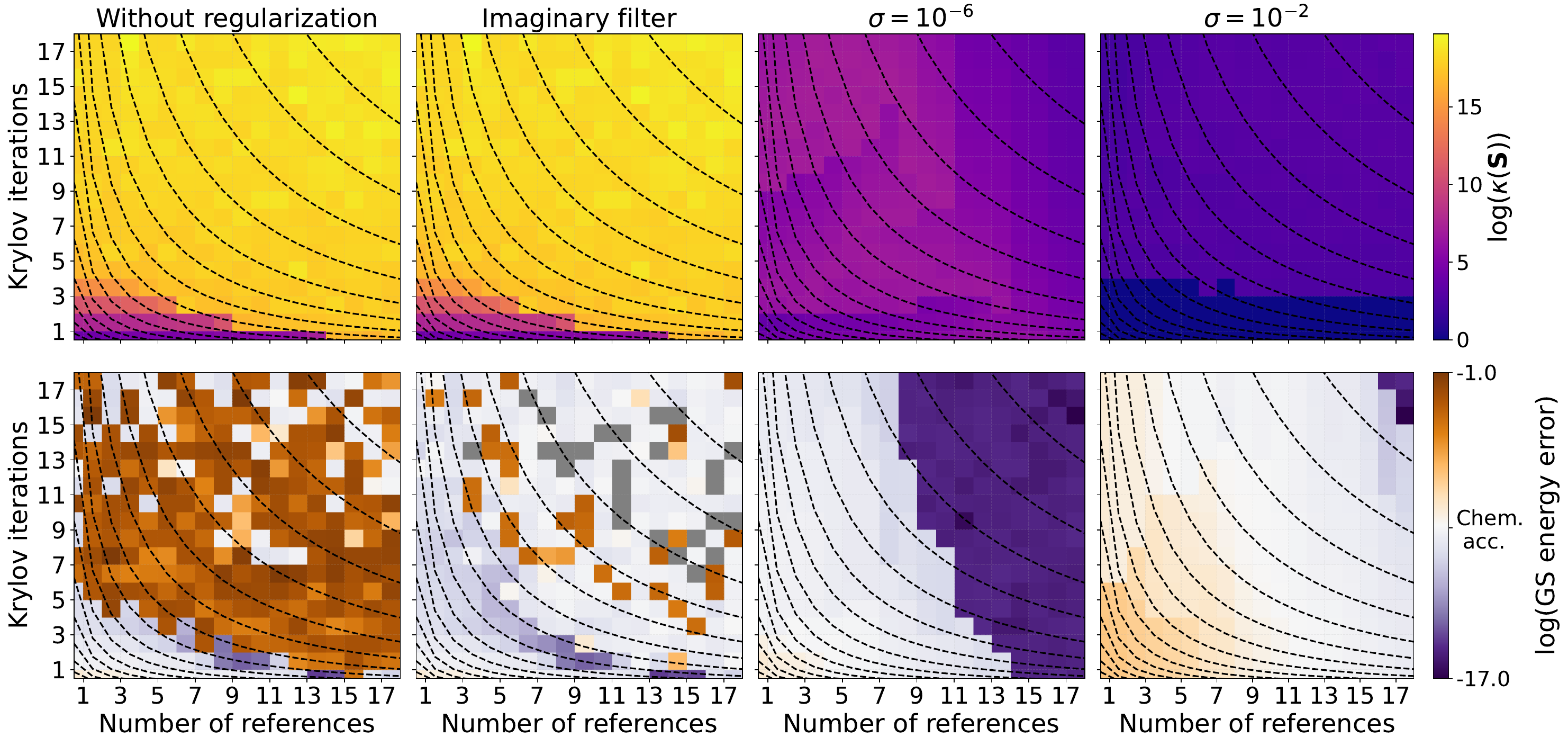}
        \caption{\textbf{QBKS-H algorithm convergence for the \ce{BeH2} triangle starting from different numbers of reference states and for different numbers of Krylov iterations.}
        The top panel shows the logarithm of the condition number, whereas the bottom panel shows the logarithm of the absolute GS energy error with regard to the exact eigenvalue. Note that the black dashed lines correspond to points with equivalent cost, i.e., the same number of distinct quantum circuits that are required to be sampled. The gray squares correspond to results where either the regularization procedure or the unitary filtering eliminated all the solutions.
        }
        \label{fig:size_beh2H}
    \end{figure}

\begin{figure}
        \centering
        \includegraphics[width=1\linewidth]{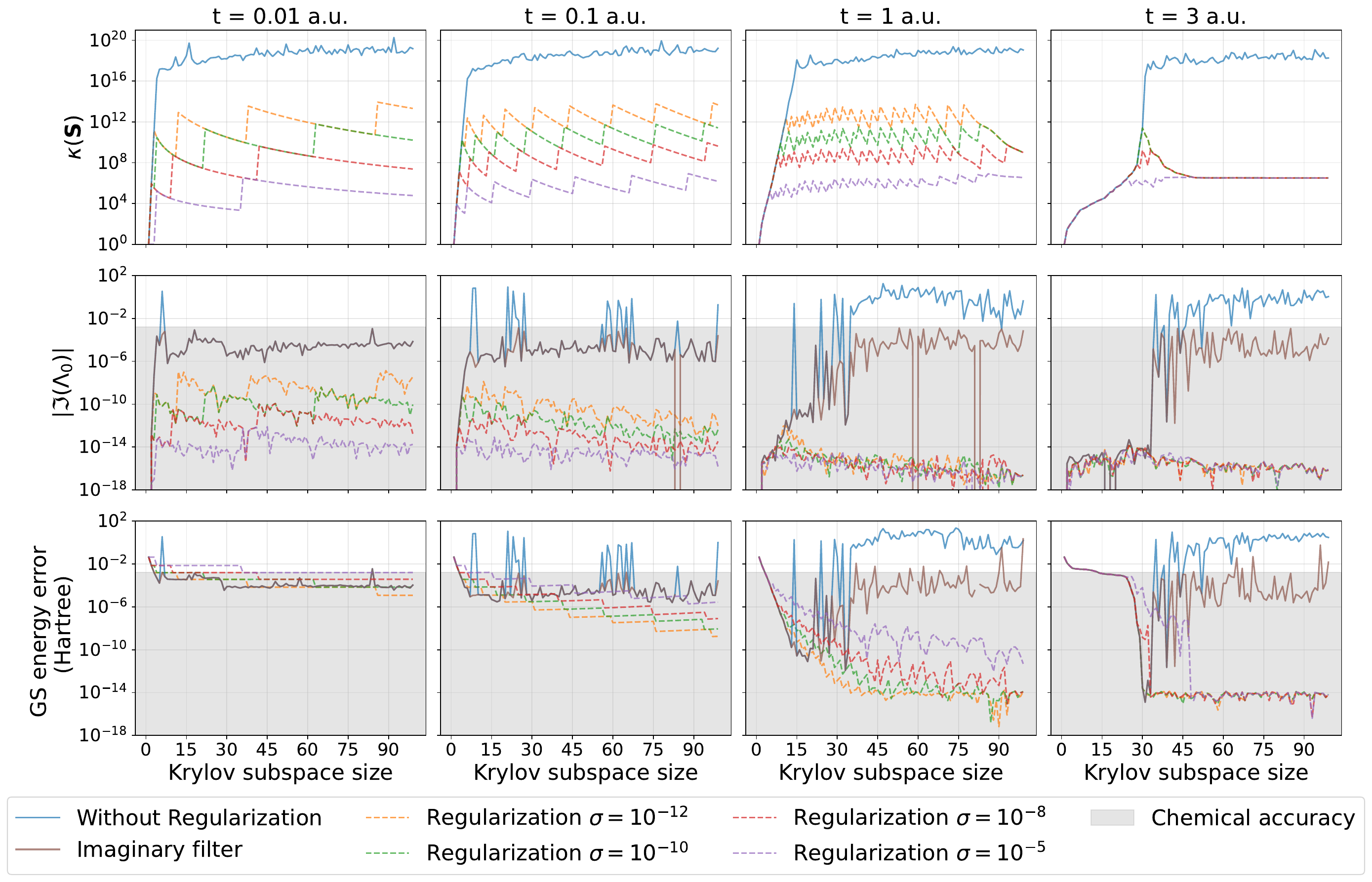}
        \caption{\textbf{QKS-H algorithm convergence for the \ce{BeH2} molecule using subspaces generated with different evolution times and several regularization values.} The top panel shows the condition number of the overlap matrix in each Krylov iteration. The middle panel shows the error in the norm of the GS energy as a function of the Krylov iteration, whereas the bottom panel shows the absolute ground energy error with regard to the exact eigenvalue. 
        }
        \label{fig:time_beh2H}
    \end{figure}

Regarding the QBKS-H algorithm, in \autoref{fig:size_beh2H} the Krylov subspace growth analysis is shown, and the time step analysis is displayed in \autoref{fig:time_beh2H}. Both figures are analogous to the ones in the main document, and result in similar conclusions to the ones presented in the main text.

\FloatBarrier

\section{\ce{H6} rectangle supplementary results}
\begin{figure}
        \centering
        \includegraphics[width=1\linewidth]{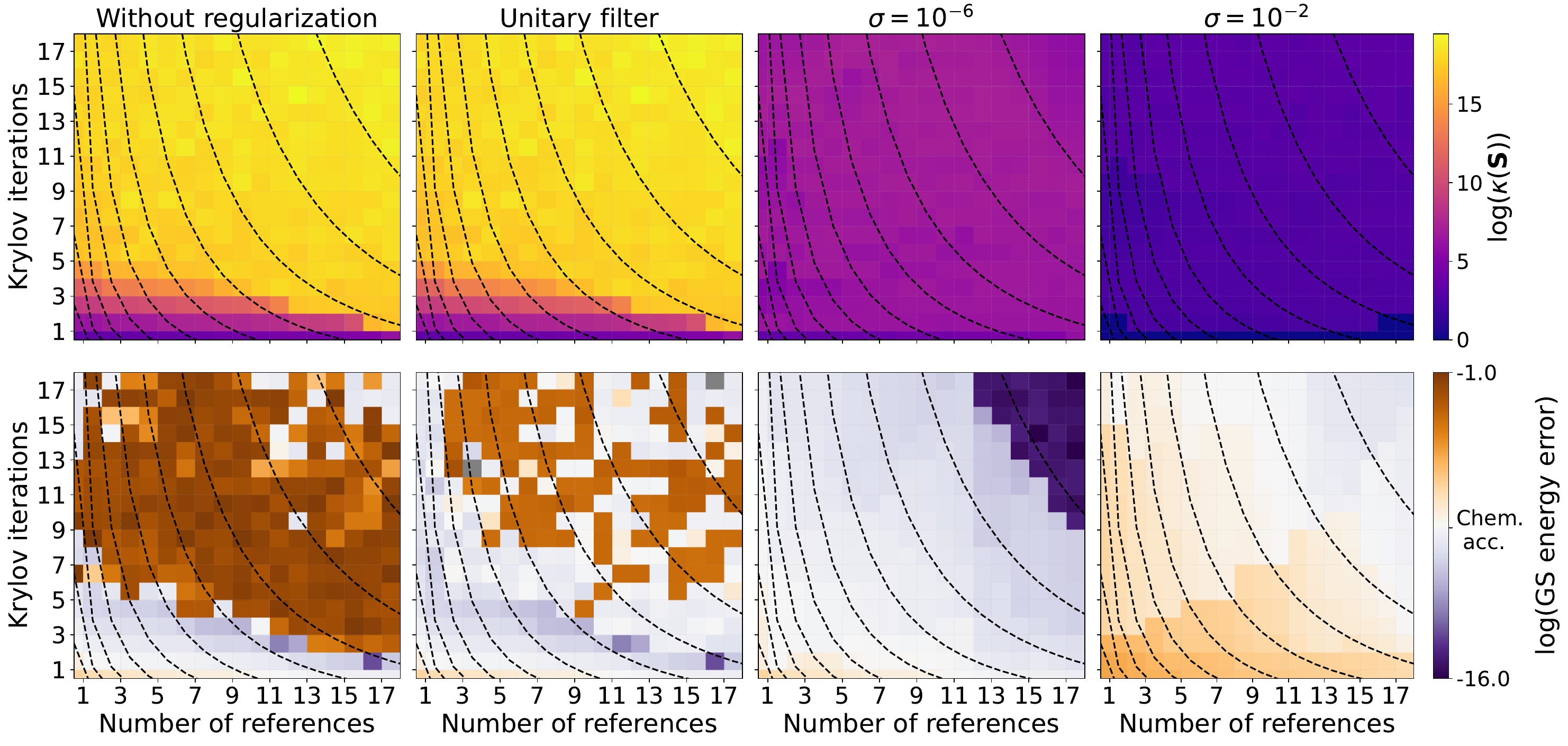}
        \caption{\textbf{QBKS-U algorithm convergence for the \ce{H6} rectangle starting from different numbers of reference states and for different numbers of Krylov iterations.} The top panel shows the logarithm of the condition number, whereas the bottom panel shows the logarithm of the absolute GS energy error with regard to the exact eigenvalue. Note that the black dashed lines correspond to points with equivalent cost, i.e., the same number of distinct quantum circuits that are required to be sampled. The gray squares correspond to results where either the regularization procedure or the unitary filtering eliminated all the solutions.
        }
        \label{fig:size_H6U}
    \end{figure}

\begin{figure}
        \centering
        \includegraphics[width=1\linewidth]{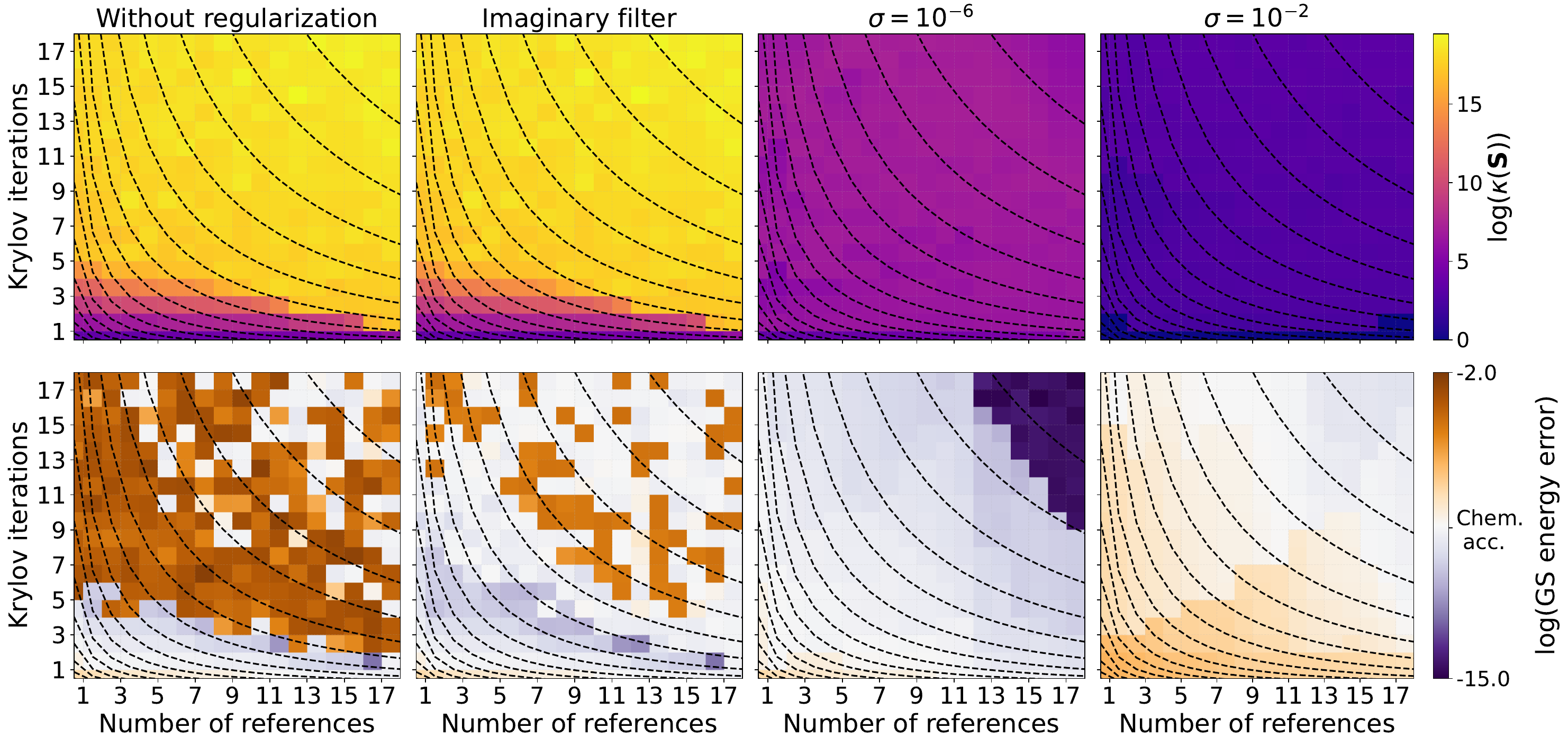}
        \caption{\textbf{QBKS-H algorithm convergence for the \ce{H6} rectangle starting from different numbers of reference states and for different numbers of Krylov iterations.} The top panel shows the logarithm of the condition number, whereas the bottom panel shows the logarithm of the absolute GS energy error with regard to the exact eigenvalue. Note that the black dashed lines correspond to points with equivalent cost, i.e., the same number of distinct quantum circuits that are required to be sampled. The gray squares correspond to results where either the regularization procedure or the unitary filtering eliminated all the solutions.
        }
        \label{fig:size_H6H}
    \end{figure}

\begin{figure}
        \centering
        \includegraphics[width=1\linewidth]{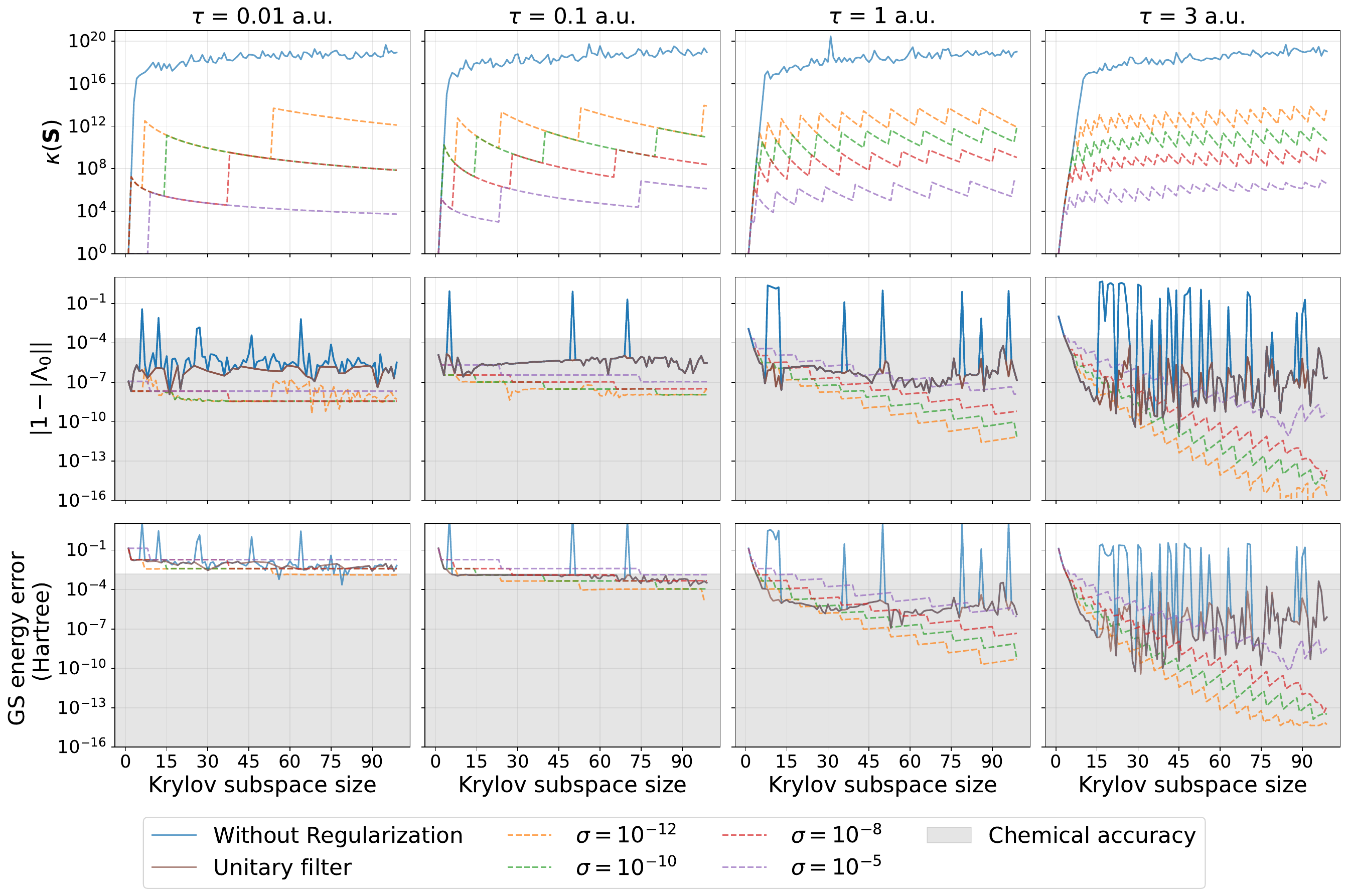}
        \caption{\textbf{QKS-U algorithm convergence for the \ce{H6} rectangle using subspaces generated with different evolution times and several regularization values.} The top panel shows the condition number of the overlap matrix in each Krylov iteration. The middle panel shows the error in the norm of the GS energy as a function of the Krylov iteration, whereas the bottom panel shows the absolute ground energy error with regard to the exact eigenvalue. 
        }
        \label{fig:time_H6U}
    \end{figure}

\begin{figure}
        \centering
        \includegraphics[width=1\linewidth]{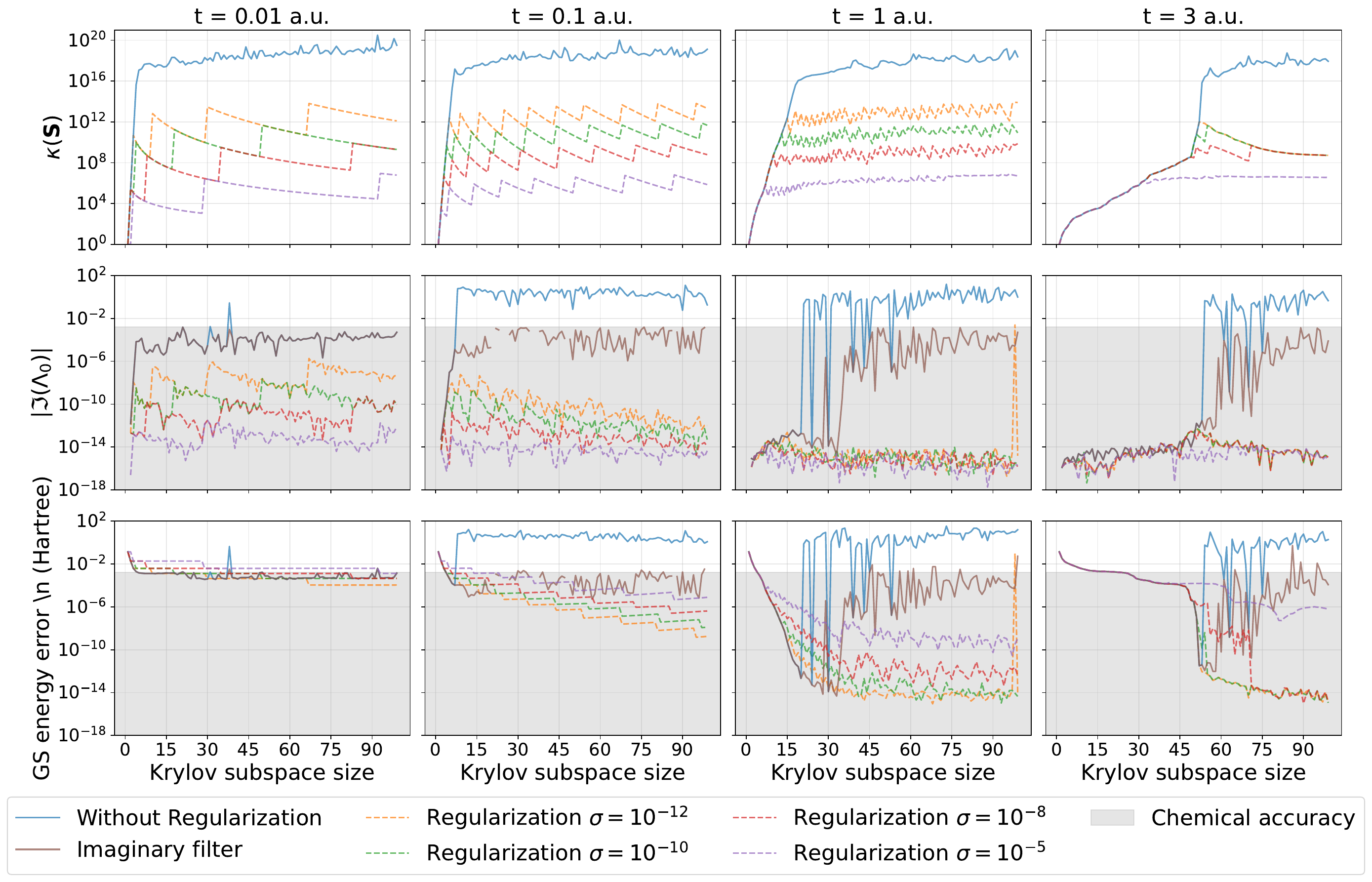}
        \caption{\textbf{QKS-H algorithm convergence for the \ce{H6} rectangle using subspaces generated with different evolution times and several regularization values.} The top panel shows the condition number of the overlap matrix in each Krylov iteration. The middle panel shows the error in the norm of the GS energy as a function of the Krylov iteration, whereas the bottom panel shows the absolute ground energy error with regard to the exact eigenvalue. 
        }
        \label{fig:time_H6H}
    \end{figure}

This section contains the supplementary Q(B)KS-U and -H algorithms analysis for the \ce{H6} rectangle system. In \autoref{fig:size_H6U} and \autoref{fig:size_H6H} we show the Krylov subspace growth analysis and in \autoref{fig:time_H6U} and \autoref{fig:time_H6H} the time step analyses are shown for both algorithms.
The conclusions drawn from the results of this test system are the same as the ones concluded for \ce{BeH} as presented in the main document.

\FloatBarrier

\section{Singular value distribution or $\mathbf{S}$}
\label{app:singular}
\begin{figure}
    \centering
    \includegraphics[width=1\linewidth]{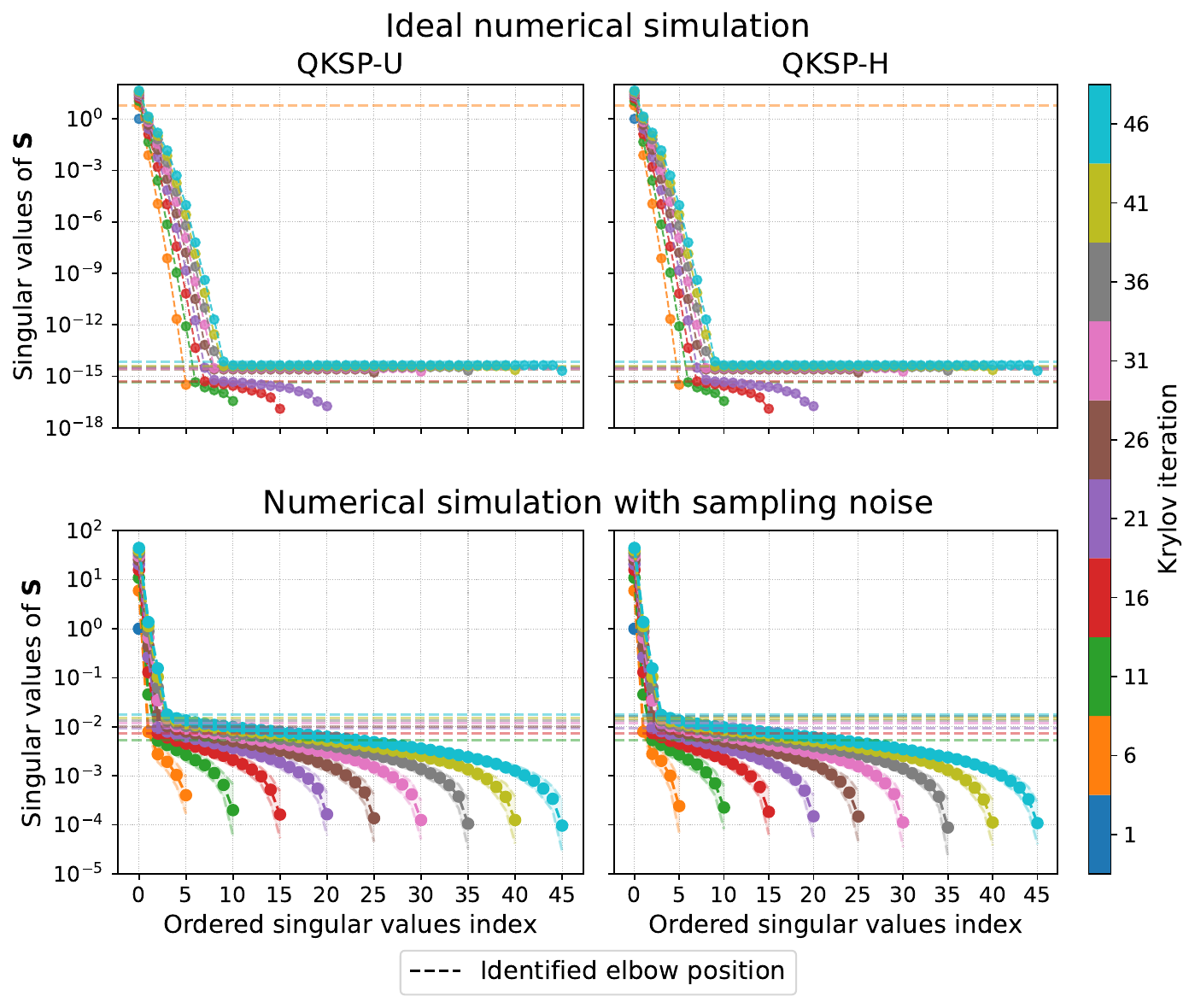}
    \caption{\textbf{Unregularized singular values of $\mathbf{S}$ for \ce{BeH2} for both QKS-U and QKS-H algorithms.} 
    The time-evolution step for the QKS-U is $1$ a.u. and for QKS-H is $\frac{1}{||\hat{H}||}$ a.u., for the fairness of the results, since QKS-U requires normalization of the results and -H does not.
    For the shot noise simulations, the geometric mean and standard deviation were calculated from 100 independent executions, as indicated by the shaded areas.
    The horizontal dashed lines correspond to the geometric mean of the identified elbow position.}
    \label{fig:singular}
\end{figure}

For further clarification and intuition on the elbow regularization method, in \autoref{fig:singular} we show here the $\log_{10}$ distribution of the singular values of $\mathbf{S}$ for noiseless and shot noise simulations of both QKS-U and QKS-H.
The singular values in both cases follow a L-shaped distribution. 
This motivated the use of a regularization method where the singular values smaller than the elbow should be discarded.
Note that the spread in singular values obtained with independent shot noise simulations is minimal, suggesting that the singular value distribution of $\mathbf{S}$ is relatively stable even in the presence of statistical noise.
